\documentclass[9pt,twocolumn,twoside,amsmath,amssymb,lineno,prl]{revtex4-1}

\usepackage{verbatim}
\usepackage{graphicx}
\usepackage{bbold} 

\usepackage[final]{changes}

\begin{document}

\title{Enhancing the predictability and retrodictability of stochastic processes}

\author{Nathaniel Rupprecht, Dervis Can Vural}
\email{Corresponding Author: dvural@nd.edu}
\affiliation{University of Notre Dame}

\begin{abstract}
Scientific inference involves obtaining the unknown properties or behavior of a system in the light of what is known, typically, without changing the system. Here we propose an alternative to this approach: a system can be modified in a targeted way, preferably by a small amount, so that its properties and behavior can be inferred more successfully. For the sake of concreteness we focus on inferring the future and past of Markov processes and illustrate our method on two classes of processes: diffusion on random spatial networks, and thermalizing quantum systems.
\end{abstract}

\maketitle

Much of science revolves around inference; reconstructing the unknown from what is known \cite{anderson1958introduction,le1990maximum,box2011bayesian}. Observable patterns here and now inform us of inaccessible patterns out and away. We can reconstruct the history of life from available fossils \cite{turner2000functions,slater2012integrating,gavryushkina2014bayesian}, or predict the fate of the universe by observing the present night sky \cite{krauss2000life,ulanowicz2009increasing,frautschi1982entropy}; we can infer hidden states and transition probabilities \cite{baum1966statistical,nasrabadi2007pattern,fine1998hierarchical}, connections and weights of neural networks \cite{boyen1998tractable,stevenson2008inferring,nguyen2017inverse} or parameters, initial states and interaction structures of complex systems \cite{ghonge2017,besag1974spatial,cocco2007reconstructing,iba2008inference,gomez2010inferring,lenglet2004inferring,haas2013expectation,ghahramani1996parameter,lokhov2014inferring,altarelli2014patient}.

Ordinarily, inference is a passive, non-disruptive process. \added{Unlike engineering, natural sciences are motivated by knowing nature, rather than changing it.} However knowing and changing are not necessarily mutually exclusive. Earlier, it was established that attempting to describe and predict a system can inadvertently influence it, potentially even rendering it indescribable and unpredictable \cite{vural2011, rupprecht2017}. Here we study the converse case of how the intrinsic properties of a system can be purposefully modified so that its past or future is more \emph{inferable}.

A number of authors addressed the problem of predicting the future and retrodicting the past of a stochastic process  \cite{crutchfield2009time,ellison2009prediction,tatem2006global,rupprecht2018limits}. In this study, we are concerned not with finding specific strategies or algorithms to optimally predict or retrodict stochastic systems, but rather finding ways to optimally \emph{modify} stochastic systems so that they become more predictabile or retrodictable. \added{An engineer might use control theory to balance a bipedal robot, stabilize the turbulent flows surrounding a wing, or maximize the signal to noise ratio in an electric circuit \cite{control}. Here we do the same, but optimize the susceptibility of a system to the inquiry of its past and future.} 

Forward in time, the entropy associated with the probability distribution of the system state will increase monotonically, as per $H-$theorems \cite{carnevale1981h,ramshaw1993h,shiino2001free}. A similar trend also holds backwards in time \cite{rupprecht2018limits}. Thus, our task is to determine how transition rates should be perturbed infinitesimally as to minimize the generation of inferential entropy in either temporal direction. After establishing a general theoretical framework, we implement these ideas to two specific example systems. The first is a diffusion process taking place on a spatial random network. The second is a quantum harmonic oscillator with a time-dependent temperature.

\section{Quantifying Predictability and Retrodictability}
Concentrated, sharply peaked probability distributions informing the future or past states of a stochastic system \added{are closer to being deterministic than higher entropy distributions}. Accordingly, we use the Gibbs-Shannon entropy to measure the degree of inferrability \cite{shannon2001mathematical}, and later on show that this indeed is a good measure. Given a stochastic process, \(X_t\), characterized by its transition matrix \(T_\alpha(\omega) = Pr(X_t = \omega \vert X_0 = \alpha)\), and initial state $\alpha$, the entropy of the process at a final time \(t\) is
\begin{align}
S_T(\alpha) = -\sum_\omega T_\alpha(\omega) \log T_\alpha(\omega).
\label{PredictionEntropy}
\end{align}
When \(X_t\) is the state of a thermodynamic system, this is the standard thermodynamic entropy. In the present information theoretical context we refer to $S_T$ as the ``prediction entropy.''

Naturally, the \added{average} entropy generated by a process  depends on how it is initialized - the prior distribution \(P^{(0)}\). To characterize the the process itself, we marginalize over the initial state, \(\alpha\),
\(
\langle S_T \rangle = \sum_\alpha P^{(0)}(\alpha) S_T(\alpha),
\)
where $P^{(0)}(\alpha)$ is the probability of starting at $\alpha$. 
Likewise, we quantify the retrodictability of a process by a ``retrodiction entropy'', $\langle S_R \rangle = \sum_\omega P^{(t)}(\omega) S_R(\omega)
$. Here, $R_\omega(\alpha)$ is the probability the system started in state $\alpha$ given that the observed final state was \(\omega\), \(S_R\) is its entropy analogous to (\ref{PredictionEntropy}), and \(P^{(t)}(\omega)\) is the probability that the process is in state \(\omega\) at time \(t\) unconditioned on its initial state. 

Interestingly, the predictability and retrodictability of a system are tightly connected: Since $S_T$ and $S_R$ are related by the Bayes' theorem,  $R_\omega(\alpha) = T_\alpha(\omega) P^{(0)}(\alpha)/\sum_{\alpha^\prime} T_{\alpha^\prime}(\omega) P^{(0)}(\alpha^\prime)$, it follows that $S_T$ and $S_R$ are also related \cite{rupprecht2018limits},
\begin{align}
\langle S_R \rangle = \langle S_T \rangle - (S_t - S_0)
\label{CentralTheorem}
\end{align}
where \(S_0\) is the entropy of the prior probability distribution $P^{(0)}$, and \(S_t\) is the entropy of \(P^{(t)}(\omega) = \sum_\alpha P^{(0)}(\alpha)\,T_\alpha(\omega)\). 

We use $\langle S_T \rangle$ and $\langle S_R \rangle$ to measure how well we can predict the future and retrodict the past of a stochastic process. The higher the entropies, the less certain the inference will be.

\section{Variations of Markov Processes}
\added{In a Markov process, the state of a system fully determines its transition probability to other states. Markov processes accurately describe a number of phenomena ranging from molecular collisions through migrating species to epidemic spreads \cite{lemons2002introduction,prinz2011markov,urban2005modeling,black2012stochastic,rohlf2008reactive}.}

Consider a Markov process defined by a transition matrix \(T_{ji}\), which we view as a weighted network. 

A system initialized in state $i$ with probability \( P^{(0)}_i\), upon evolving for $t$ steps, will follow a new distribution $P^{(t)}_i = \sum_j P^{(0)}_j (T^t)_{ji}$.
Accordingly,
\begin{align}
\langle S_R \rangle = &-\sum_{i,j} P^{(t)}_j (R^t)_{ji} \log (R^t)_{ji}
\nonumber\\
\langle S_T \rangle = &-\sum_{i,j} P^{(0)}_j (T^t)_{ji} \log (T^t)_{ji}.
\end{align}
Thus both entropies depend on the duration of the process \(t\). Note that probability is normalized \( \sum_i (T^t)_{ji} = 1 \) for all \(j\), \(t\).

Suppose that it is somehow possible to change the physical parameters of a system slightly, so that the probability of transitions are perturbed, \(T_{ji} \rightarrow T_{ji}^\prime = T_{ji} + \epsilon\,q_{ji}\), where \(\epsilon\) is a small parameter. For now, we do not assume any structure on $q$, other than implicitly demanding that it retains probabilities within $[0,1]$ and preserves the normalization of rows. This variation leads to a change in the \(t\)-step transition matrix,
\begin{align}
(T+\epsilon\,q)^t &= T^t + \sum_{p=1}^t \epsilon^p \eta^{(t,p)}
\label{MultiStepPerturbation} \\
\eta^{(t,p)} &= \sum_{1\le k_1 < ... < k_p \le t} T^{k_1-1} \xi_{k_2-k_1-1} \xi_{k_3-k_2-1}\ldots \xi_{t-k_p}\nonumber
\end{align}
where \(\xi_k = q\,T^k\). The superscripts of \(\eta^{(t,p)}\) refer to the power of the transition matrix, \(t\), and the order of the contribution, \(p\), which is analogous to the order of the derivative of a function. So \(\eta^{(t,p)}\) is the \(p\)-th order contribution to the varied \(t\)-step transition matrix. This defines a set of \(p\)th order effects for the \(n\)th power of the transition matrix. In the sequel, we will be studying first variations, therefore, we will only need
\begin{align}
\eta^{(t,1)}\equiv\eta^{(t)} = q T^{t-1}+T q T^{t-2} + \dots + T^{t-1} q.
\label{FirstOrder}
\end{align}

The difference between the the entropies of the perturbed and the original systems is $
\Delta \langle S_{T,R} \rangle \equiv \langle S_{T,R}(T^\prime)\rangle - \langle S_{T,R}(T) \rangle.
$ Whenever \(\Delta \langle S_{T,R} \rangle\) is of order \(\epsilon\) and higher, we can evaluate the \emph{variation}
\begin{align}
\delta \langle S_{T,R} \rangle = \lim_{\epsilon \to 0} \Delta \langle S_{T,R} \rangle/\epsilon,\label{var}
\end{align}
which in essence is the derivative of \(\langle S_T \rangle\) or \(\langle S_R \rangle\) in the \(q\) ``direction.''

With little algebra, we can show that the first order perturbations of the \(t\)-step entropies \(\langle S_R \rangle\) and \(\langle S_T \rangle\) are
\begin{align}
\Delta \langle S_T \rangle &= -\epsilon \log \epsilon \sum_{i,j} \mathbb{1}_T^c P^{(0)}_j \eta_{ji}^{(t)} 
\label{PerturbedST} \\
& -\epsilon \sum_{i,j} P^{(0)}_j \eta_{ji}^{(t)} \left[ \mathbb{1}_T (1+\log (T^t)_{ji}) + \mathbb{1}_T^c \log \eta_{ji}^{(t)} \right]\nonumber\\
\Delta \langle S_R \rangle &= -\epsilon \log \epsilon \sum_{i,j} \mathbb{1}_T^c P^{(0)}_j \eta^{(t)}_{ji}\\
\!\! -\epsilon \sum_{i,j} &P^{(0)}_j \eta^{(t)}_{ji} \left[ \log [(T^t)_{ji}/P^{(t)}_i] + \mathbb{1}_T^c \left( \log [\eta^{(t)}_{ji}/P^{(t)}_i] - 1\right) \right]\nonumber
\end{align}
The Kronecker functions \(\mathbb{1}_T\) and \(\mathbb{1}_T^c\) which implicitly depend on the indices \(i,\,j\), and the time, \(t\), are defined to be \(\mathbb{1}_T=0\) if \((T^t)_{ji} = 0\) and to equal 1 otherwise, and \(\mathbb{1}_T^c = 1-\mathbb{1}_T\). 

As we see, the $\epsilon\log\epsilon$ terms can cause the limit (\ref{var}) to diverge, causing a sharp, singular change in entropy generation. This is expected. The divergence will happen only when the perturbation enables a  path between two states where there was none. This is because \((T^t)_{ji}=0\) only if \(i\) could not be reached from \(j\) in \(t\) steps, but if this is still true after perturbation, the \(\eta_{ji}\) term will be zero. 

On the other hand, if the perturbation does not enable a path between two isolated states, but preserves the topology of the transition matrix, then (\ref{PerturbedST}) simplifies considerably; the divergent \(\mathbb{1}^c\) terms vanish, and we take the limit,
\begin{align}
& \delta \langle S_T \rangle = -\sum_{i,j} P^{(0)}_j \eta_{ji}^{(t)} [1+\log (T^t)_{ji}]\nonumber\\
& \delta \langle S_R \rangle = - \sum_{i,j} P^{(0)}_j \eta^{(t)}_{ji} \log [(T^t)_{ji}/P^{(t)}_i]\label{SimplerPerturbedSR}
\end{align}

Having established a very general theoretical framework, we now implement these ideas on two broad classes of stochastic systems for which the structure of the perturbing matrix \(q\) is specified further. We first consider random transition matrices drawn from a matrix ensemble. Secondly, we study a physical application - we enhance the predictability and retrodictability of thermalizing quantum mechanical systems by means of an external potential.

\subsection{Improving the predictability and retrodictability of Markov processes}  

\begin{figure}
\centering
\includegraphics[width=0.45\textwidth]{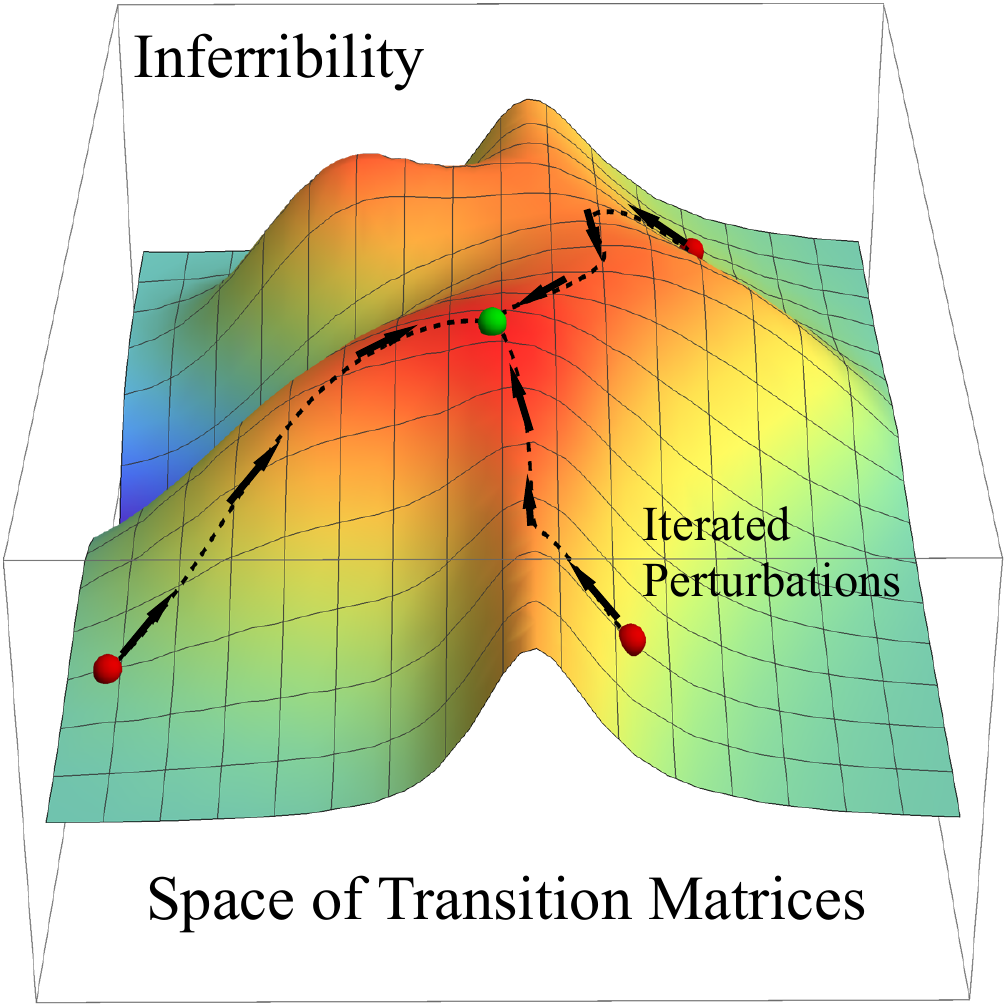}
\caption{\textbf{Ascending the space of transition matrices to maximize predictability and retrodictability.} Each point in the space of Markov transition matrices, represented by the x,y plane has an associated predictive and retrodictive entropy. Equation (\ref{GradientSR}) allow us to find the direction in network space - parameterized by the transition rates \(T_{ji}\) - in which entropy locally increases (or decreases) the most. Perturbations can then be applied to move the network in that direction, leading to a system that is more susceptible to inference. Red dots represent different starting networks which climb along the black paths, via gradient ascent, to an entropy maximum, represented by a green dot.}
\label{Sketch}
\end{figure}

\begin{figure*}
\includegraphics[width=\textwidth]{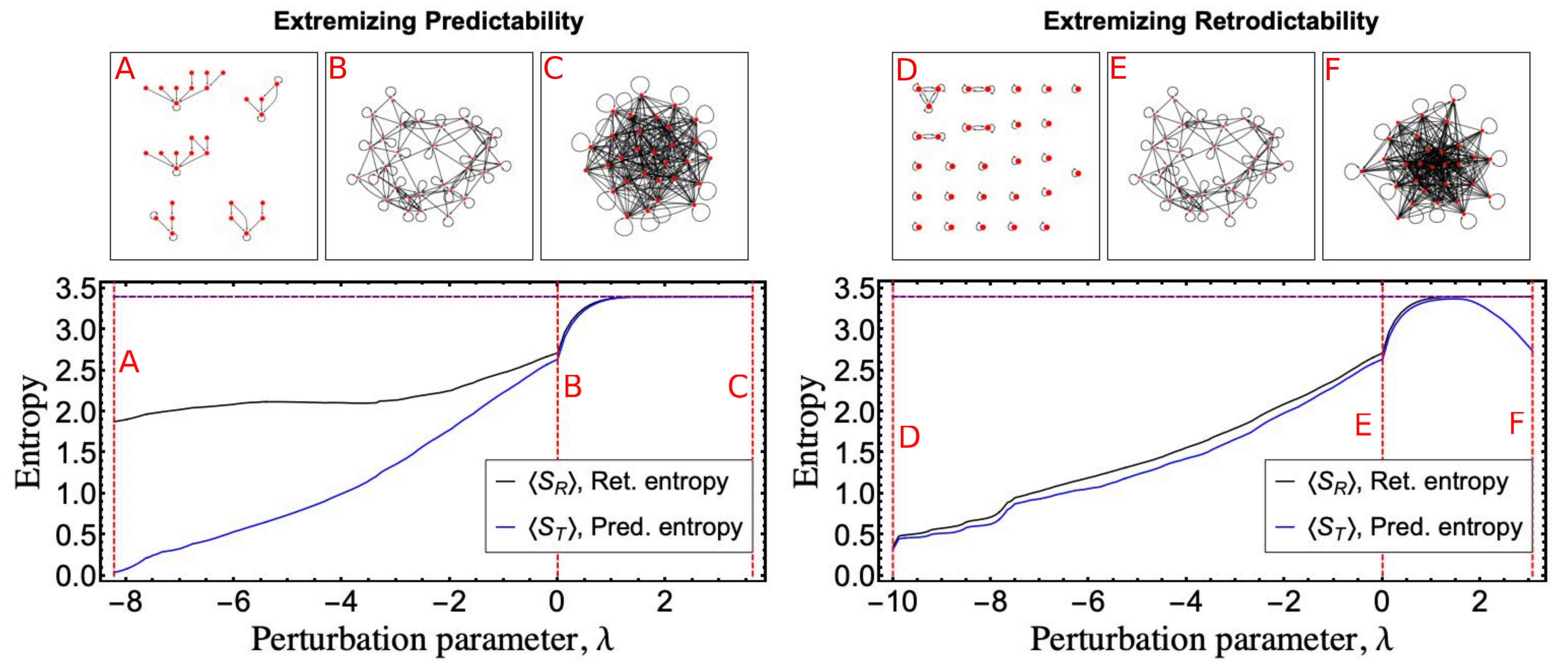}
\caption{\textbf{Entropy extremization of a Markov Process.} Entropy during the evolution of a Markov network according to the extremization procedures (\ref{GradientSR}). The parameter \(\lambda\) corresponds to ``how many times'' the perturbation operator has been applied, or how far along the gradient curve, \(\nabla S\), we have pushed the transition network. The graphs in panels (A-F) are pictorial representations of the Markov transition matrices. The points in the evolution that we sample graphs from are marked with red lines and a letter. We choose to look at the minimal entropy graph, initial graph (\(\lambda=0\)), and maximal entropy graph both for prediction entropy and retrodiction entropy. The entropy curves, \(\langle S_T \rangle\) and \(\langle S_R \rangle\) correspond to how easy it is to predict the final state or retrodict the initial state of the Markov process. The network is a geometric network as described above, with \(\beta = 0.5\) and \(n=30\) states. We optimize our entropies for a t=3 step process. \added{The purple line along the top marks the maximum possible entropy.} \textbf{Left:} How the entropies change as we extremize \(\langle S_T \rangle\), and three samples of transition probability networks. \textbf{Right:} How the entropies change as we extremize \(\langle S_R \rangle\), and three samples of transition probability networks.}
\label{Fig:MatrixEvolution}
\end{figure*}

We start by studying a general class of perturbations that can be applied to an arbitrary Markov process, and evaluate the associated entropy gradient, which can be thought as the direction in matrix space that locally changes \(\langle S_R \rangle\) or \(\langle S_T\rangle\) the most (Fig. \ref{Sketch}). As we climb up or down the entropy gradient, we show how the transition matrix evolves (Fig. \ref{Fig:MatrixEvolution}). 

We consider \added{a family of} perturbations that vary the relative strength of any transition rate. This involves changing one element in the transition matrix while reallocating the difference to the remaining nonzero rates so that the total probability remains normalized. In other words,
\begin{align}
\Delta_{\beta \alpha}^{(\epsilon)} T_{ji} = T_{ji} + \epsilon \cdot \mathbb{1}_{j \beta} (\mathbb{1}_{i \alpha} - T_{\beta i}).
\label{MatrixPertubation}
\end{align}
To first order in \(\epsilon\), this is the same as adding \(\epsilon\) to the \((i,j)\) element, and then dividing the row by \(1+\epsilon\) to normalize, so it is a natural choice for a perturbation operator. It also obeys \(\Delta^{(\epsilon)}\Delta^{(-\epsilon)} T = \mathbb{1} + \mathcal{O}(\epsilon^2)\). We define the perturbation acting on a zero element to be zero if \(\epsilon<0\) since elements of the transition matrix must be non-negative. From (\ref{MatrixPertubation}) and (\ref{MultiStepPerturbation}), we obtain the perturbed matrices and perturbed \(\langle S_R \rangle\).

\added{To study the effect of many perturbations of the form (\ref{MatrixPertubation}), applied successively, we carry out a gradient ascent algorithm in matrix space. At each point, we change the transition rates infinitesimally to maximally increase or decrease retrodiction or prediction entropy. We parameterize the gradient ascent by \(L^2\) distance in matrix space, i.e. \(\mathrm{d}(A, B) = \|A-B\| = \left[ \sum_{i,j} (A_{ji}-B_{ji})^2 \right]^{1/2}\).}

\added{In a gradient descent algorithm, one descends over a function \(f(r)\) over a parameter \(t\) (time) by solving \( \dot{\vec{r}} = \nabla f(\vec{r}) / \| \nabla f(\vec{r}) \| \), where the normalization ensures that \(\| \partial_t \vec{r}(t) \| = 1\), so the total distance of the path \(\vec{r}(t)\) just \(t\).}

\added{Similarly, we define our gradients in entropy space to be either \(\Delta_{ji} \langle S_R(T) \rangle \) or \(\Delta_{ji} \langle S_T(T) \rangle \), depending on whether we are optimizing retrodiction or prediction. We parameterize our path, \(T(\lambda)\), so that the total distance of the path (in \(L^2\) matrix space) is \(\lambda\),
\begin{align}
\dot{T}_{ji}(\lambda) = \lim_{\epsilon \to 0} \Delta^{(\epsilon)}_{ji} \langle S_{R,T}(T_{ji}) \rangle / \| \Delta^{(\epsilon)}_{ji} \langle S_{R,T}(T_{ji}) \rangle \|.
\label{GradientSR}
\end{align}
Since we carry out this scheme numerically, using a finite difference method, it does not matter if the limit in (\ref{GradientSR}) exists. In these cases, our numerical scheme returns large, finite jumps.}

To illustrate our formalism in action, we solve (\ref{GradientSR}) for a particular \added{example system}: diffusion taking place on a directed spatial random network. \added{We build a spatial network, such that neighboring} nodes are placed at regular intervals on a circle, and are also cross connected with probability $P(S_{ji}=1) = e^{-\beta d_{ij}}$, that decay with distance $d_{ij}$ \cite{barthelemy2011spatial}. The transition matrix \(T\) is obtained by normalizing the rows of \(S\). 

An example is shown in Fig. \ref{Fig:MatrixEvolution} where the 3-step (\(t=3\)) predictability and retrodictability change as the transition matrix is perturbed iteratively. We use  perturbation operators that extremize retrodictability (left column) or predictability (right column). The transition matrices are displayed as networks, shown at the \added{maximal, zero, and minimal} $\lambda$ values, corresponding to minimal entropy state, initial state and maximal entropy state. For clarity's sake, we display only edges with weights above \(0.025\).

We now interpret our results to ensure that our theoretical framework makes qualitative sense and works as expected. First, we observe that  perturbations that maximize both $\langle S_T \rangle$ and $\langle S_R \rangle$ displace the transition matrix towards the same point: in both cases $T$ evolves to a point where \((T^t)_{ji} = p_i\), a probability vector. In other words the probability of transition does not depend on what state the system is currently in. Taking the \(3^{\text{rd}}\) power of the extremal \(T\) reveals that this is indeed the case, although, of course, \(T\) itself can retain some complex structure (see Fig. \ref{Fig:MatrixEvolution}C, F). As expected, when a system moves from any state to any other state with equal likelihood, it is most difficult to infer its past or future.

In contrast, \emph{minimizing} entropy produces two very different transition matrices\added{, for \(\lambda \ll 0\),} depending on the type of entropy we minimize. Minimizing retrodiction entropy tends to eliminate branches and fragmenting the network into linear chains \added{(including isolated nodes) as seen in \ref{Fig:MatrixEvolution}D}. Probability flows through these unidirectionally, thus retrodiction involves nothing more than tracing back a linear path.

On the other hand, the global minima of the prediction entropy yields transition matrices in which \added{all probability in each connected component flows towards a single node, reachable in \(t\) steps. A process where the initial state uniquely determines the final state is indeed trivial to predict. This type of network can be seen in Fig.\ref{Fig:MatrixEvolution}A, where there are five connected components, each with a single accumulating node}.

This also explains why \(\langle S_R \rangle\) tends to stay the same in the \(\lambda<0\) direction when minimizing \( \langle S_T \rangle\): If $S_t = \langle S_T \rangle = 0$, then (\ref{CentralTheorem}) implies \(\langle S_R \rangle = S_0\), which is the maximum possible value for \(\langle S_R \rangle\). This can also be understood intuitively - when a final measurement is made, the system is always found to be in a unique accumulating state, and this yields no information about what state the system started in. If, however, the minimal \(\langle S_T \rangle\) network instead \added{has multiple connected components and collector nodes}, \(\{k_j\}\), then there can be a decrease in \(\langle S_R \rangle\) since \(R_{k_0}\),  \(R_{k_1}\), ... are different distributions. 

Observing the entropy curves for both figures, there are obvious non-differentiable points in the entropies at \(\lambda = 0\). This is because there are sites that are not connected to one another after the requisite number of time steps, so the \(\epsilon \log \epsilon\) terms in (\ref{PerturbedST}) are non-zero. It is also apparent that there are other non-differentiable points for \(\lambda < 0\). These are due to matrix elements vanishing, essentially hitting the boundary of the simplex that the elements can travel in (the space of possible matrix elements for each row is a simplex because of the conditions \(T_{ji}>0\) and \(\sum_i T_{ji} = 1\)). 

Lastly, we observe in the figures that \(\langle S_T \rangle \le \langle S_R \rangle\). This follows from (\ref{CentralTheorem}). \(S_0\) is maximal for our prior, a uniform distribution, so \(S_t - S_0 \le 0\). 

\begin{figure}[h]
\includegraphics[width=0.5\textwidth]{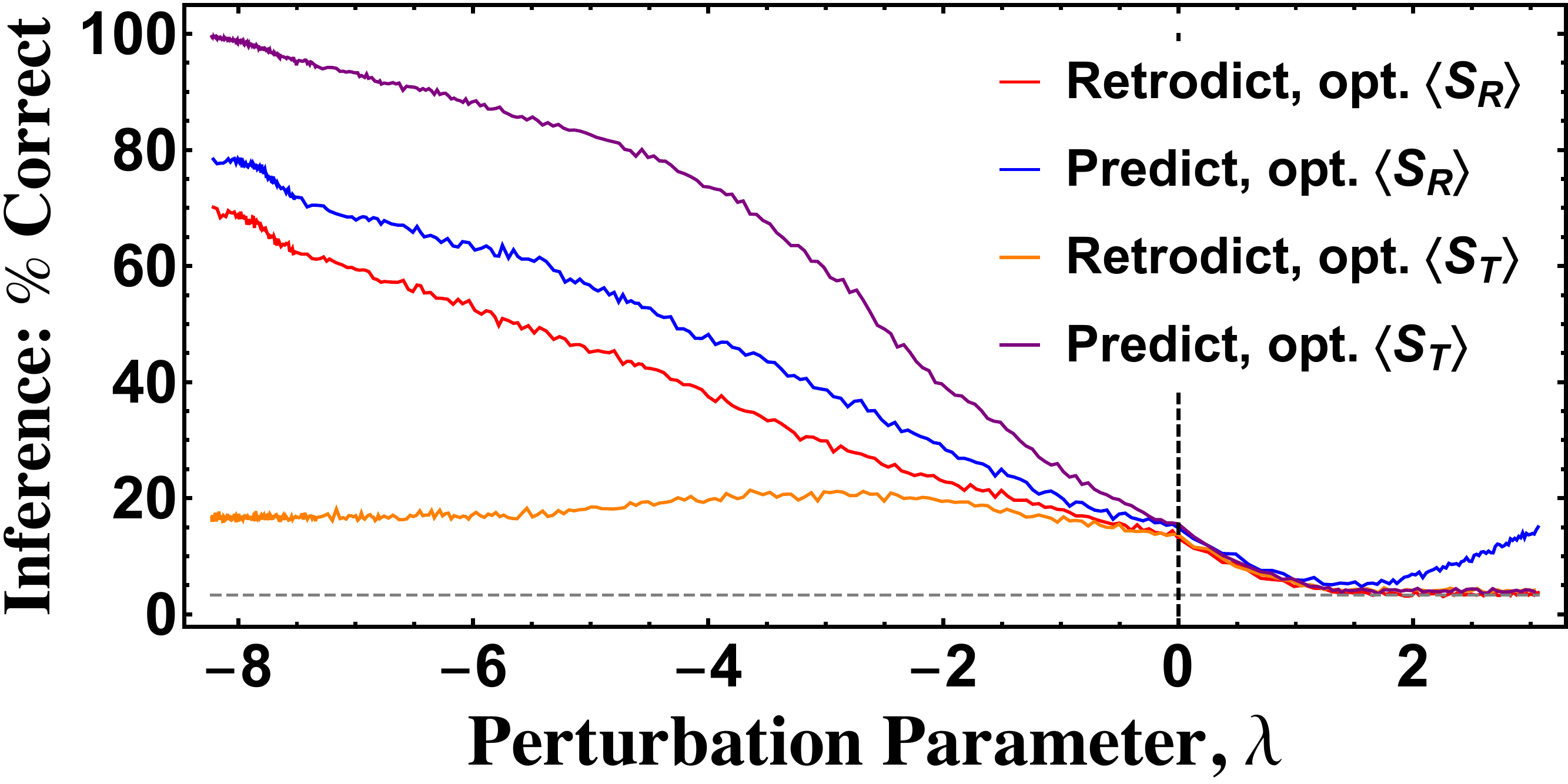}
\caption{\textbf{Performance in predicting initial or final states.} The performance of prediction and retrodiction on the evolving Markov transition networks - the same evolutions as shown in Fig. \ref{Fig:MatrixEvolution}. The four cases plotted are either correct inferences of the \emph{initial} state (retrodiction) or correct inferences of the \emph{final} state (prediction) while either optimizing \(\langle S_R \rangle\) or optimizing \(\langle S_T \rangle\). As a baseline, making random guesses, the strategy would obtain the initial or final state correctly 3.3\% of the time (since there are 30 states). This baseline is depicted as a dashed gray line.}
\label{Fig:Improvement}
\end{figure}

So far, we have only extremized entropy, but have not shown that this leads to a significant difference in our ability to infer the past or future. To do so, we must run stochastic processes, predict or retrodict final or initial states, and report how often our inference is successful. For systems with very large number of states, the probability mass per state will be very small, \added{thus a suitable metric of success may be the rate at which our $n$ best guesses are correct, for an $n$ that scales with system size. Instead, we adopt a harder metric, and simply report how often we infer the correct state, spot on.}

For predictions we pick an initial state, \(i\), with a uniform prior, evolve it forward stochastically, and then make our inference by picking the state with the highest probability \(T_{ji}\) and checking if we are correct. Retrodictions are carried out similarly, except we observe the outcome of the stochastic simulation, \(j\), and infer the initial state by picking the state with highest probability \(R_{ji}\). For both inferences we counted how often our estimation agreed, and ran the stochastic simulation and inference many times.

The result can be seen in Fig. \ref{Fig:Improvement}. The transition matrices we did our test is the same as those shown in Fig. \ref{Fig:MatrixEvolution}. The success rate of predicting final states and retrodicting initial states while optimizing either \(\langle S_R \rangle\) or \(\langle S_T \rangle\) is plotted. Since there are 30 states in our network, the baseline accuracy is \(1/30 = 3.3\%\), as marked with a dashed gray line. Our success rate aligns nicely with the entropy in Fig. \ref{Fig:MatrixEvolution}, left column. We reach almost \(100\%\) accuracy when we minimize \(\langle S_T \rangle\) (Fig. \ref{Fig:Improvement}, purple line), which corresponds to the point where \(\langle S_T \rangle\) nearly vanishes.

The improvement in retrodictability always lags behind predictability. This is because \(\langle S_R \rangle\) must be greater than \(\langle S_T \rangle\), as per Eq. (\ref{CentralTheorem}). 

\added{Naturally, descending an entropy landscape all the way returns transition matrices with trivial structure and dynamics. In our diffusion example, one could have guessed from the beginning, that a network with only inward branches, or one with disconnected chains, would be much more predictable than an all-to-all network with equally distributed weights. However, our formulation is useful not because it eventually transforms every network into a trivial network, but because it provides the steepest direction \emph{towards} a trivial network. Secondly, our formulation is also useful because among many trivial networks, it moves us towards the direction of the \emph{closest} one. Thus, we must really ask how effective \emph{small} pertubations are, far before the system turns into a trivial one.}

\added{We find indeed, that significant differences to inferential success can be made with relatively small changes to the transition matrix. Table \ref{Tab:ChangesTable} quantifies how much the transition matrix has been modified, versus how much our retrodictive (top three rows) and predictive (bottom three rows) success have improved. For example, the fifth row shows that if we would like to be spot-on correct in predicting the final state of a stochastic process with $30$ states and $900$ transitions, our success rate can be improved by \(\sim5\%\) by modifying only $8$ out of $900$ transition rates by more than $0.1$, with none being larger than $0.2$.  The cumulative change in all transition rates for this perturbation totals to $4.34$. The changes required to improve our success rate by $10\%$ are not much larger.}


\added{As a final point of interest, we see that for all the \(\pm 5\%\) in Table \ref{Tab:ChangesTable}, \(\lambda\) and the \(L^2\) distance are almost identical. This means that to get from the initial matrix to the perturbed matrices, one could follow the gradient calculated at the initial matrix in a straight line - the path is roughly straight in matrix space for at least that distance.}

\begin{table}
 \begin{tabular}{| c c c c c c c |} 
 \hline
 \(\lambda\) & \text{Optimize} & \(\Delta \%\) & \(L^2 \text{ dist.}\) & \(L^1 \text{ dist.}\) &  \(\Delta \ge 0.1\) & \(\Delta_{max}\) \\ [0.5ex] 
 \hline\hline
 +0.49 & \(\langle S_R \rangle\) & -5.29\% & 0.49 & 8.96 & 4 & 0.117 \\ 
 \hline
 -1.18 & \(\langle S_R \rangle\) & +5.12\% & 1.03 & 9.69 & 39 & 0.23 \\
 \hline
 -2.13 & \(\langle S_R \rangle\) & +10.2\% & 1.63 & 16.17 & 86 & 0.31 \\
 \hline
 \hline
 +0.49 & \(\langle S_T \rangle\) & -5.72\% & 0.49 & 8.81 & 4 & 0.134 \\
 \hline
 -0.556 & \(\langle S_T \rangle\) & +4.92\% & 0.54 & 4.34 & 8 & 0.165 \\
 \hline
 -1.03 & \(\langle S_T \rangle\) & +10.3\% & 0.91 & 7.81 & 23 & 0.222 \\
 \hline
\end{tabular}
\caption{\added{\textbf{Matrix retrodictability and structure.} A summary of how much the initial matrix must be changed to vary the retrodiction success. The columns are perturbation parameter (\(\lambda\)), what type of entropy was optimized (retrodiction or prediction, ``optimize''), change in performance (retrodiction or prediction, corresponding to what was optimized) from the original matrix (\(\Delta \%\)), the \(L^2\) distance from the original matrix, the \(L^1\) distance from the original matrix, the number of transition probabilities that were changed more than \(0.1\) (\(\Delta \ge 0.1\)), and the maximum change of any transition probability. The original matrix had \(126\) nonzero entries and had retrodictability/predictability successes of \(13.6\%\) and \(15.1\%\). There are 900 transition that can be modified.}
}
\label{Tab:ChangesTable}
\end{table}

\subsection{Improving inference for a physical system through external fields}
In a physically realistic scenario, it is unlikely to have full control over individual transitions. An experimentalist can only tune physical parameters, such as external fields or temperature, which influence the transition matrix indirectly. Furthermore, it is often not practical to vary physical parameters by arbitrarily large amounts. Thus ideally we should improve predictability and retrodictability optimally, while only applying small fields.

To meet these goals, we consider a class of quantum systems in or out of equilibrium with a thermal bath. These systems are fully characterized by eigenstates $\psi_1,...,\psi_n$ with energies \(E_1,...,E_n\) undergoing Metropolis-Hastings dynamics \cite{metropolis1953equation} where a system attempts to transition to an energy level above or below with equal probability; an attempt to decay always succeeds, while an attempt to excite succeeds with probability \(\exp[-\beta (E_{k+1} - E_k)]\). 
\begin{align}
T_{k,j} = \begin{cases} 
      \frac{1}{2} \exp[-\beta(E_{k+1}-E_k)] & j=k+1 \\
      \frac{1}{2}( 1-\exp[-\beta(E_{k+1}-E_k)]) & j=k \\
      \frac{1}{2} & j=k-1 \\
      0 & |j-k|>1
   \end{cases}
   \label{tmatrix}
\end{align}
Furthermore we assume that the ground state \(E_0\) cannot decay, and the highest state \(E_n\) is unexcitable. For the regime of validity of Markovian descriptions of thermalized quantum systems, we refer to \cite{gardiner2004quantum,kapral2006progress}.

We now determine the effects of a small perturbing potential \(v(x)\). The perturbation will shift the energy levels, which changes the transition matrix, which in turn changes the average prediction and retrodiction entropies of the system. Our goal is to identify what perturbing potential would maximally change these entropies. Since we are concerned with the first order variation in entropy, it will suffice to also use first order perturbation theory to calculate energy shifts.

The perturbed \(k\)-th energy level is \(E_k = E_k^{(0)} + \epsilon \cdot \delta E_k\). When the perturbation is applied the exponential terms in \(T\) change as
\begin{align*}
& \mathrm{e}^{-\beta (E_{k+1} - E_{k})} \rightarrow 
 \mathrm{e}^{-\beta (E_{k+1} - E_{k}) -\epsilon\, \beta (\delta E_{k+1} - \delta E_{k}) }
\\
& = \left[1-\epsilon\,\beta (\delta E_k - \delta E_{k-1}) \right] e^{-\beta (E_k - E_{k-1})} + \mathcal{O}(\epsilon^2).
\end{align*}
From this, we can find our first order change \(T_{ji}^\prime = T_{ji} + \epsilon \, q_{ji}\) in terms of the change in energy levels, \(\delta E_k\),
\begin{align}
q_{kj} &= -\beta (\delta E_{k+1} - \delta E_{k}) \exp[-\beta(E_{k+1}-E_{k})] \cdot S_{kj} 
\nonumber \\
S_{kj} &= \mathbb{1}_{j,k+1} - \mathbb{1}_{j,k} = \begin{cases} 
      +1 & j=k+1 \\
      -1 & j=k \\
      0 & j\ne k, j\ne k+1 
   \end{cases}.
\label{QuantumPerturbationMatrix}
\end{align}

Now we will write the prediction and retrodiction entropy \(\delta \langle S_{T,R} \rangle\) variations as a functional of a perturbing potential, and then use calculus of variations to obtain the extremizing potential. For clarity, we will derive our equations in one dimensions; the generalization to higher dimensions is straightforward.

We partition the spatial domain, \(\Omega\), into \(N\) intervals, \([x_i, x_{i+1})\), of width \(\Delta x\) and let our potential be a piecewise constant function of the form
$
v(x) = \sum_{i=0}^{N-1} v_i\, \mathbb{1}_{x \in [x_i, x_{i+1})}.
$
As \(N \to \infty\), the first order change in the \(k\)-th energy level is
\begin{align}
\delta E_k = \langle \psi_k \vert v \vert \psi_k \rangle \sim \sum_{i=0}^{N-1} v_i |\psi(x_i)|^2 \Delta x 
\label{dEEquation}
\end{align}
\added{since \(\int_{x_i}^{x_{i+1}} v_i \vert \psi(x) \vert^2 \sim v_i \vert \psi(x_i) \vert^2 \Delta x \).}
We substitute the \(\delta E\)s, (\ref{dEEquation}), into (\ref{QuantumPerturbationMatrix}) to get the \(q\) matrix,
\begin{align}
& q_{k j} = \sum_{i=0}^{N-1} v_i\,\beta\, \left[|\psi_{k}(x_i)|^2 - |\psi_{k+1}(x_i)|^2 \right] e^{-\beta(E_{k+1}-E_k)}S_{k j} \Delta x \nonumber\\
&\equiv \sum_{i=0}^{N-1} v_i\,q_{k j}(x_i)\, \Delta x\to\int_\Omega v(x) \tilde{q}_{kj}(x)\mathbf{d}x
\label{QAtX}
\\
& \added{\tilde{q}_{kj}(x) \equiv \beta \left(\vert \psi_k(x)\vert^2 - \vert \psi_{k+1}(x) \vert^2 \right) \, e^{\beta(E_{k+1} - E_k)} \cdot S_{kj}}
\end{align}
we substitute this in
into (\ref{FirstOrder}) to get
\begin{align*}
& \eta^{(t)}_{ji} = \int_\Omega \mathbf{d}x \, v(x) \sum_{k=0}^t T^k \tilde{q}(x) T^{t-k-1} \equiv \int_\Omega \mathbf{d}x \, v(x)\,\tilde{\eta}^{(t)}_{ji}(x)
\\
& \added{\tilde{\eta}^{(t)}_{ji}(x) \equiv \sum_{k=0}^{t} T^k \tilde{q}(x) \, T^{t-k-1}}
\end{align*}
and therefore,
\begin{align}
\delta \langle S_{T,R} \rangle[v] &= -\int_\Omega \mathbf{d}x \, v(x) \frac{\delta^2 \langle S_{T,R} \rangle}{\delta x\,\delta v}
\end{align}
where \(\delta^2 \langle S_{T,R} \rangle / \delta x\,\delta v\) is (\ref{SimplerPerturbedSR}) with \(\tilde{\eta}^{(t)}_{ji}(x)\) substituted in for \(\eta^{(t)}_{ji}\).

Lastly, we ensure the smallness of the perturbation by introducing a penalty functional, \(C[v] = \frac{1}{2} \gamma \int v(x)^2 \mathbf{d}x\) and ask what potential \(v(x)\) extremizes
\begin{align}
\mathsf{F}_{T,R}= \delta \langle S_{T,R} \rangle - C=\int_\Omega \left(v(x) \frac{\delta^2 \langle S_{T,R} \rangle}{\delta x \, \delta v} - \frac{1}{2} \gamma \,v(x)^2 \right) \mathbf{d}x.\nonumber
\end{align}
We take a variational derivative with respect to \(v(x)\) and set it to zero to obtain the extremizing potential,
\begin{equation}
v_{T,R}(x) = -\frac{1}{\gamma} \frac{\delta^2 \langle S_{T,R} \rangle}{\delta x \, \delta v}.
\label{ExtremalSolution}
\end{equation}
\added{This \(v_{T,R}\) is the external potential that extremizes the gradient of entropy minus the penalty functional.}

\begin{figure*}
\includegraphics[width=\textwidth]{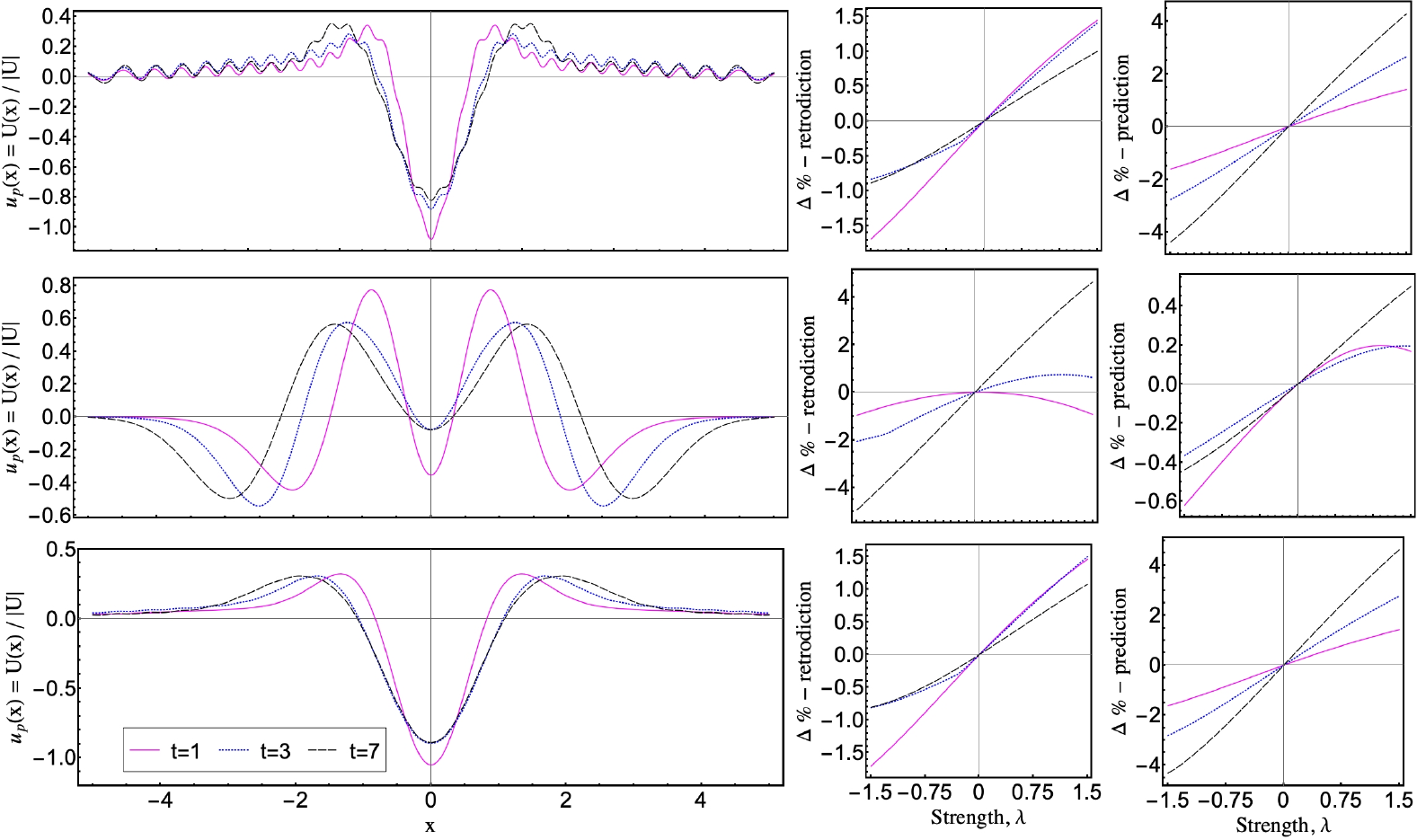} 
\caption{\textbf{The external fields and performance checks.} We take \(\hbar=\omega=m=1\) and plot perturbations that minimize  \(\langle S_R \rangle\) or \(\langle S_T \rangle\) for processes taking $t=1,3,7$ time steps (\ref{ExtremalSolution}). \added{Along with the perturbing potential, we plot the change in inference success (how accurately we can guess the initial or final state of the system) vs. strength of the applied perturbation. Locally, this curve should have positive slope.} \textbf{Top Row:} A high temperature ($T=10$) equilibrium system is quenched to a low temperature ($T=1$) system. These potentials extremize \(\langle S_R \rangle\). \textbf{Middle Row:} A low temperature ($T=1$) system quenched to a high temperature ($T=10$) system. Note the large scale shape of the potential is similar to that of the top panel. These potentials extremize both \(\langle S_R \rangle\) and \(\langle S_T \rangle\). \textbf{Bottom Row:} A high temperature ($T=10$) equilibrium system is quenched to a low temperature ($T=1$) system. These potentials extremize $\langle S_T \rangle$.}
\label{PotentialPerturbations}
\end{figure*}

\subsection{Improving the predictability and retrodictability of a thermalizing quantum harmonic oscillator} 
We can now ask what perturbing external field should be applied a quantum harmonic oscillator that is in the process of warming up or cooling down, in order to improve its predictability or retrodictability. For this system \(V(x) = \frac{1}{2} m \omega^2 x^2\), and \(E_k = (k+\frac{1}{2}) \hbar \omega \). The stationary eigenfunctions are \(\psi_k(x) = \frac{1}{\sqrt{2^k k!}} \pi^{-1/4} \exp \left( - \frac{x^2}{2} \right) H_k(x) \) where \(H_k\) is the \(k\)-th Hermite polynomial, \(H_k(x) = (-1)^k e^{x^2} \frac{d^k}{d x^k} e^{-x^2}\). For concreteness, we also have to choose a prior distribution on states. We choose the prior distribution to be an equilibrium distribution at a (possibly different) temperature, \(P_k \propto e^{-\beta_2 E_k}\). We truncate the transition matrix at an energy $E_n\gg 1/\beta_1, 1/\beta_2$ so that edge effects are negligible. We take \(m=\hbar=\omega=1\)\added{, and choose \(U\) to be the negative of (\ref{ExtremalSolution}) so that adding them to \(V(x)\) \emph{decreases} the corresponding entropy, and \emph{increases} inference performance.}

\added{The initial and final temperatures determine the flow of probability. The equilibrium distribution at a high temperature has much more probability mass at higher energy states than an equilibrium distribution at a low temperature, so if we start with a high temperature and quench to a low temperature, there will tend to be a flow of probability from high states to low states. The opposite will happen when we quench from a low to a high temperature.}

\added{To ensure that each perturbing potential, \(U(x)\), actually increase or decrease predictability/retrodictability (depending on whether we add or subtract it from \(V(x)\)), we calculate the  ``\(\Delta\,\%\)'' for retrodiction and prediction: the percent difference in how often we can correctly guess the initial or final state, upon perturbing the system. The performance is obtained similarly to that in Fig. \ref{Fig:Improvement} (cf. methods section). The perturbation potential is normalized to \(u_p(x) = U(x)/ \|U\|\) so that the \(L^2\) norm of $u_p(x)$ is \(1\), and the strength, \(\lambda\), with which \(u_p\) is applied is varied so that the total potential is \(V(x) + \lambda\,u_p(x)\).}

Fig. \ref{PotentialPerturbations} shows some extremizing potentials for a system that was at one temperature, and is then suddenly quenched to a different temperature. Potentials that extremize prediction and retrodiction entropy for different number of steps forward and backward in time are shown. \added{Alongside the potential we plot the change in inference success, as the potential is applied at varying strengths.} For a high initial temperature and low final temperature, the extremal perturbation adds small ripples to the original one (Fig. \ref{PotentialPerturbations}, top and bottom rows). In Fig. \ref{PotentialPerturbations}, middle row, the system is initially at a lower temperature, and then is quenched to a high temperature. This perturbation potential happens to extremize both \(\langle S_R \rangle\) and \(\langle S_T \rangle\). 

\added{To quantify how significantly the perturbations change the quantum system, we keep track of the \(L^1\) difference in eigenvalue spacing, i.e. \(\mathcal{S} \equiv \sum_{k=1}^{30} \vert E^\prime_k - E^\prime_{k-1} - \hbar \omega \vert\).
The largest values \(\mathcal{S}\) achieves for any potential and applied strength shown in Fig. (\ref{PotentialPerturbations}) is \(\sim 1.2\). In other words, we can get few percent change in success rate by introducing a change to all energy levels that amounts to one level spacing. Note that this is a single step perturbation along a single direction, rather than an iterated one.}

This example illustrates how to combine real, physical, continuous quantities, such as perturbation potentials, with the more abstract formalism of evaluating the entropy of Markov transition matrices with discrete states. The general procedure we outlined in this section can also be applied to other thermal systems, quantum or otherwise. 


\section{Discussion}
We developed a formalism to describe exactly how predictability and retrodictability changes in response to small changes in a transition matrix, and used it to descend entropy landscapes to optimally improve the accuracy with which the past or future of a stochastic system can be inferred. Our main results are the equations relating perturbations of Markov processes to the change in average entropy and retrodiction entropy of the system, (\ref{MultiStepPerturbation}, \ref{PerturbedST}, \ref{SimplerPerturbedSR}). 

Our formulas lead us to intuitive results such as the divergence of entropy generation when a path between two otherwise isolated states is enabled. However, they also lead us to less obvious conclusions, such as how predictability changes when retrodictability is optimized (and vice versa); or the shape of optimal potentials perturbing a thermalizing quantum system.

Our basic equations, (\ref{MultiStepPerturbation}, \ref{PerturbedST}, \ref{SimplerPerturbedSR}) are very generally applicable to any discrete-time Markov process. The type of transition matrix perturbations we chose to study, namely (\ref{MatrixPertubation}) and  (\ref{QuantumPerturbationMatrix}) are natural and practical choices, but of course, they are not the only two possibilities. For example, an operator that takes two matrix elements \(0 < T_{j a}, T_{j b} < 1\) and ``transfers'' probability between them, changing them to \(T_{j a} + \epsilon, T_{j b} - \epsilon\) would make an interesting future study.

In our work we observed an intriguing asymmetry between prediction and retrodiction. In particular, we observe that predictability is more easily improved than retrodictability. This a byproduct of how we set up our problem: We took the initial distribution, \(P^{(0)}\), and the forward dynamics, \(T\), as givens, and found the probability, \(P^{(t)}\), via propagating \(P^{(0)}\) with \(T\). If we had done the opposite by picking the distribution $P^{(t)}$ and the backwards dynamics, \(\tilde{T}\), then we could find $P^{(0)}$ to be the back-evolved distribution, then our results would reverse. 

An experimenter only has control over the prior distribution at the current time, \(P^{(0)}\), but cannot in general decide what distribution she wants at an arbitrary future time, \(P^{(t)}\), and pick a \(P^{(0)}\) that results in a specified \(P^{(t)}\). The fact that we set up the problem so that \(t=0\) was the ``controlled'' time, and the state at the final time is the \emph{result} of the choices made at \(t=0\) ultimately lead to the seeming emergence of an ``arrow of time'' \cite{arrow}.

\added{Since our method makes changes to a system to extremize the average of a function over a set of trajectories, it could well be considered within the domain of stochastic control theory \cite{aastrom2012introduction,forte2017iterative}. However, there are various elements in our approach that depart from classical stochastic control, which typically deals with problems of the form
\begin{align*}
    dX_t &= f\left(X_t, v(t) ; t\right) + \hat{\xi}_t
    \\
    C(X_0, v) &= \left\langle \phi(X_T) + \int_0^T R(X_t, v; t) \, \mathbf{d}t \right\rangle_{P(\xi)}
\end{align*}
where \(X_t\) is the system trajectory, \(\hat{\xi}\) is a Weiner process, \(v\) is a control parameter, \(C\) is a cost function, and \(\phi\) and \(R\) are the target cost and some function that quantifies cost-of-control,  cost-of-space, cost-of-dynamics, etc \cite{chernyak2013stochastic}. The goal is to find the \(\tilde{v}\) that minimizes \(C\).}

\added{One difference is that we do not restrict ourselves to a Weiner process, but allow any valid transition matrix. The control parameter, \(v\), could be the perturbation to the original transition matrix, or it could be some other external parameter which indirectly results in a change in the transition matrix, as in the thermalizing quantum oscillator example.}

\added{The second difference is the structure of our cost function. In our case, the cost is an average weighted over priors. For prediction entropy,
\begin{align*}
    \langle S_T \rangle &= \langle C(X_0) \rangle_{P^{(0)}} = -\langle \langle \log P(X_T \vert X_0) \rangle_{P(\cdot \vert X_0)} \rangle_{P^{(0)}}.
\end{align*}
For a delta function prior, this reduces to the standard control theory cost function, which depends on the initial condition of the system.
For retrodiction entropy \(S_R(X_T)\) the cost depends on the final state, and is then averaged over the posterior distribution of \(X_T\),
\begin{align*}
    \langle S_R \rangle = \langle C(X_T) \rangle_{P^{(T)}} = -\langle \log R(X_0 \vert X_T) \rangle_{R(\cdot \vert X_T)} \rangle_{P^{(T)}}.
\end{align*}
}

\added{The third difference is a philosophical one. Standard stochastic control aims to find a control protocol that is a global minimum of the cost function - one obtains the field \(v\) such that \(C[v+\delta v] = 0\) for all \(\delta v\). In contrast, we look for the \emph{variation} \(\delta v\) such that \(C[\delta v]\) is \emph{maximal}, where \(C\) is \(\langle S_R\rangle\) or \(\langle S_T\rangle\). Our method descends entropy gradients in a space of system parameters, and is only guaranteed to be optimal locally. This could then be paired with a stochastic gradient descent algorithm or simulated annealing to find optima in a larger neighborhood. 
In passing, we note that for systems with a very large number of states, it would probably be computationally advantageous to use a stochastic algorithm even to compute the local gradient.}

\added{There is still plenty of room to make our framework more useful and general. Currently, we assume constant transition rates, and perturb the transition matrix at a single instant. However, transition rates can be time-dependent, in which case we would have to perturb the transition rates differently at different times. Another interesting avenue would be to further explore the costs associated with changing the transition probabilities. Another natural generalization is to extend the problem to continuous time.}

\section{Summary}
Here we demonstrated \added{an active method of} inference by subjecting systems to small perturbations so that the accuracy in inferring their past or future changes maximally. 

We specifically focused on Markov processes, not only because it yields to mathematical analysis, but also because many important processes in physical, biological and social sciences are Markovian. That being said, the general principle outlined here can also be used in systems with memory, or in other inference problems such as the determination of unknown boundary conditions, system parameters, or driving forces.

As examples of \added{manipulating predictability and retrodictability}, we studied two specific types of perturbations, (\ref{MatrixPertubation}, \ref{QuantumPerturbationMatrix}), and used these to study how certain types of transition matrices evolve as they flow along the trajectory of maximal increase in retrodictability and predictability. We found that the transition networks tend to cull their connections and split into cycles and chains when we try to minimize retrodiction entropy. Conversely, the transition networks become fully connected when we attempt to maximize either inferential entropy. If one does not have full control over transition rates, one can steer a system towards the direction of either extreme by a small amount. Finally, as a physical example, we studied how to find the perturbing potential that extremally changes the predictability and retrodictability of a thermalizing quantum system. 
\section{Methods}

\textbf{Extremization of entropy.} We started with a random geometric graph, \(T(\lambda=0)\), from the ensemble described in the text, where nodes \(i\) and \(j\) are connected with probability \(e^{-\beta\,d(i,j)}\). We used \(n=30\) node graphs, with \(\beta=0.5\). The extremization is done numerically and iteratively, as outlined in (\ref{GradientSR}). 
The entropy was the entropy for a \(t=3\) step process, and we use a perturbation size \(\epsilon=0.05\), and step size \(d\lambda = 0.05\). 

At each step, the matrix of change in entropy (per \(\epsilon\)) due to perturbation of an element is calculated, \(S_{ji} = \frac{1}{\epsilon} \Delta_{ji}^{(\epsilon)} \langle S[T(\lambda)] \rangle\), where the \(S\) in the angled brackets is whichever entropy we seek to extremize \(T\) over - either \(S_R\) or \(S_T\). To get the updated transition matrix, the \(j,i\) element of \(T\) is perturbed using the standard perturbation operator (\ref{MatrixPertubation}) and strength \(\epsilon^\prime = d \lambda/ \| S_{ji} \|\). The order that we apply these operators is irrelevant up to order \((\epsilon^\prime)^2\). The updated transition matrix is then the result of applying all the perturbation operators, one for each element of \(T\). At each step, the prediction and retrodiction entropy of the Markov process were calculated and saved, along with the actual matrix \(T_{ji}(\lambda)\), for plotting purposes. The change in \(\lambda\) at each step is just the \(L^2\) distance between the previous matrix and the new, perturbed matrix.

\textbf{Inference performance.} The numerical technique used to generate Fig. \ref{Fig:Improvement} is described in the paper, we would only add that each point is the result of using 10,000 trials.

\added{The inference performance can also be calculated analytically as long as we have the transition matrix, \(T_{ji}\), and the prior, \(P^{(0)}\). Since we are guessing that the maximally likely state is the correct, the formulas for this are}
\begin{align}
\added{
    C_T = \sum_j P^{(0)}_j \text{max}_i (T^t)_{ji} 
}
    \nonumber \\
\added{
    C_R = \sum_j P^{(t)}_j \text{max}_i (R^t)_{ji}.
}
    \label{Methods:InferencePerformance}
\end{align}
\added{These formulas give us the expected fraction of times we correcly guess the final state given the initial state (\(C_T\)), or initial state given the final state (\(C_R\)). The expression \(\text{max}_i (T^t)_{ji}\) is the probability that you guess the the final state correctly \emph{given} that the initial state is \(i\), and the normalized sum simply averages your performance across all possible initial states. The \(C_R\) equation is analogous, simply substituting the retrodiction probability matrix for the transition matrix.}

\added{As expected, the performance obtained via random trials fits \(C_T\), \(C_R\) almost exactly since we are using a large number of trials.}

\textbf{Thermalizing quantum harmonic oscillator.} While it would be difficult to analytically solve (\ref{ExtremalSolution},\ref{SimplerPerturbedSR}) to find the extremal change in potential, it is a simple matter to calculate it analytically. 

For the harmonic oscillator \(V(x) = \frac{1}{2} m \omega x^2\), and \(E_k = (k+\frac{1}{2}) \hbar \omega \). The stationary eigenfunctions are \(\psi_k(x) = \frac{1}{\sqrt{2^k k!}} \pi^{-1/4} \exp \left( - \frac{x^2}{2} \right) H_k(x) \) where \(H_k\) is the \(k\)-th Hermite polynomial, \(H_k(x) = (-1)^k e^{x^2} \frac{d^k}{d x^k} e^{-x^2}\). As mentioned in the text, we choose the prior distribution to be an equilibrium distribution at a given temperature, \(P_k \propto e^{-\beta_2 E_k}\). Since we can only store finite vectors on a computer, we only track the first \(n=30\) energy eigenstates, \added{which is enough that the total probability mass (sum of Gibbs factors) the prior misses by truncation would only be \(<5\%\) of the total probability mass}. We take \(m=\hbar=\omega=1\) for simplicity. The perturbation matrix \(\eta(x)\) can be calculated numerically - it is a high order \added{(order 60)} polynomial in \(x\) times \(e^{-x^2}\) - and substituted into (\ref{SimplerPerturbedSR}) to get the (negative) extremal potential, \(U(x)\). \added{The potential is then normalized by the \(L^2\) norm of \(U\), \(u_p(x) = U(x)/ \|U\|\) where \(\|U\|=\left(\int_{-\infty}^\infty U(x)^2\,\mathbf{d}x \right)^{1/2}\).}

\added{
\textbf{Inference performance for the thermalizing quantum harmonic oscillator.} We solve for the energy eigenvalues of the harmonic oscillator potential plus the perturbation potentials using the method of shooting. For each \(u_p\), and for each strength, \(\gamma\), we numerically solve Schrodinger's equation for the potential \(\frac{1}{2}m \omega^2 x^2 + \gamma\,u_p(x)\) at different energies, \(E_{trial}\). We pick our shooting point to be far outside our region of interest, at \(x=15\), and evaluate whether the value of the numerical solution is positive or negative at the shooting point. Near an energy eigenvalue, the sign of \(\psi_{trial}\) will be (without loss of generality) less than 0 for energies a little below the true eigenvalue, and greater than 0 for energies a little above the true eigenvalue. We use the bisection method of root finding to approximate the energy eigenvalue with as much precision as we want. Our energy eigenvalues are correct up to \(10^{-6}\).
}

\added{
Once we have found the first \(n\) eigenvalues, we compute the transition matrix using (\ref{tmatrix}), which is determined by the final temperature, and the prior distribution on states, \(P^{(0)}_j = e^{-\beta_i E_j}/Z\), which is determined by the initial temperature (\(Z = \sum_{j=1}^n e^{-\beta_i E_j}\)). We then calculate the average percentage of times the final state can be inferred given the initial state after \(t=1,\,3,\,7\) steps. We use (\ref{Methods:InferencePerformance}) to do this.
}


%

\end{document}